\documentclass[aps,prc,twocolumn,showpacs,amsmath,amssymb,preprintnumbers]{revtex4}

\usepackage[10pt]{type1ec}  
\usepackage[T1]{fontenc}
\usepackage{CJKutf8}
\usepackage[overlap, CJK]{ruby}
\usepackage{CJKulem}
\newenvironment{SChinese}{%
  \CJKfamily{gbsn}%
  \CJKtilde
  \CJKnospace}{}

\usepackage{latexsym} 
\usepackage{graphicx}       
\usepackage{color}
\usepackage{bm}   

\newcommand{\be}{\beta}

\newcommand{\De}{\Delta}
\newcommand{\ep}{\varepsilon}

\newcommand{\la}{\lambda}

\newcommand{\beq}{\begin{equation}}
\newcommand{\eeq}{\end{equation}}
\newcommand{\ba}{\begin{array}}
\newcommand{\ea}{\end{array}}
\newcommand{\bea}{\begin{eqnarray}}
\newcommand{\eea}{\end{eqnarray}}
\newcommand{\bi}{\begin{itemize}}  
\newcommand{\ei}{\end{itemize}}
\newcommand{\ben}{\begin{enumerate}} 
\newcommand{\een}{\end{enumerate}}
\newcommand{\bc}{\begin{center}}
\newcommand{\ec}{\end{center}}

\newcommand{\dsp}{\displaystyle}
\newcommand\eqn[1]{(\ref{#1})}      

\newcommand{\ee}[1]{\times 10^{#1}}

\newcommand{\fm}{{\rm fm}}
\newcommand{\km}{{\rm km}}

\newcommand{\MeV}{{\rm MeV}}

\newcommand{\ptrans}{p_{\rm trans}}
\newcommand{\ntrans}{n_{\rm trans}}
\newcommand{\etrans}{\varepsilon_{\rm trans}}
\newcommand{\Mtrans}{M_{\rm trans}}
\newcommand{\Rtrans}{R_{\rm trans}}
\newcommand{\Msolar}{M_{\odot}}
\newcommand{\pcent}{p_{\rm cent}}
\newcommand{\ecent}{\varepsilon_{\rm cent}}

\newcommand{\cQM}{{c^{\phantom{1}}_{\rm QM}}}
\newcommand{\cNM}{{c^{\phantom{1}}_{\rm NM}}}
\newcommand{\cQMsq}{c^2_{\rm QM}}
\newcommand{\cNMsq}{c^2_{\rm NM}}
\newcommand{\destab}{\De\ep_{\rm crit}}
\newcommand{\Mmax}{M_{\rm max}}

\begin{document}

\title{Generic conditions for stable hybrid stars}
\author{Mark G. Alford, Sophia Han
(\begin{CJK}{UTF8}{}\begin{SChinese}韩 君\end{SChinese}\end{CJK})
}

\affiliation{Physics Department, Washington University,
St.~Louis, MO~63130, USA}

\author{Madappa Prakash}
\affiliation{Department of Physics and Astronomy,
Ohio University, Athens, OH~45701, USA
}

\begin{abstract}
We study the mass-radius curve of hybrid stars, assuming a
single first-order phase transition between nuclear and
quark matter, with a sharp interface between the quark matter core and nuclear
matter mantle.  
We use a  generic parameterization of the quark matter equation of state,
  which has a constant, i.e.~density-independent, speed of sound (``CSS'').
We argue that this parameterization provides a framework
for comparison and empirical testing of models of quark matter.
We obtain the phase diagram of possible forms of the hybrid star 
mass-radius relation, where the control parameters are
the transition pressure, energy density discontinuity,
and the quark matter speed of sound. We find that this diagram is sensitive
to the quark matter parameters but
fairly insensitive to details of the nuclear matter equation of state.

We calculate the maximum hybrid star mass as a function of the 
parameters of the quark matter  EoS, and find that there
are reasonable values of those parameters that give rise to
hybrid stars with mass above $2\,M_\odot$.
\end{abstract}

\date{27 July 2013} 



\maketitle

\section{Introduction}
\label{sec:intro}

There has been much work on a plausible form for the equation of state (EoS)
of nuclear matter at densities above nuclear density, using models of the
nuclear force that are constrained by existing scattering data
(see, for example, \cite{Akmal:1998cf,Gandolfi:2011xu}).
However, we remain almost completely ignorant of the quark matter EoS
at the low temperatures and the range of densities that are 
of relevance to neutron stars. This is because cold
quark matter cannot be created in laboratories, and numerical studies
using the known strong interaction Lagrangian are stymied by the
sign problem (see, for example,
\cite{Barbour:1997ej,Alford:1998sd,Cox:1999nt,Hands:2007by}).

In this work we therefore assume a generic quark matter equation
of state, and see in what ways it might be constrained by measurements
of the mass and radii of compact stars.
We assume that there is
a first-order phase transition between nuclear and quark matter,
and that the surface tension of the interface is high enough
to ensure that the transition occurs at a sharp interface (Maxwell
construction) not via a mixed phase (Gibbs construction).
This is a possible scenario, given the uncertainties in the value
of the surface tension \cite{Alford:2001zr,Palhares:2010be,Pinto:2012aq}.
(For analysis of generic equations of state that 
continuously interpolate between the phases to model mixing or percolation,
see Refs.~\cite{Macher:2004vw,Masuda:2012ed}.)

It is already known \cite{Seidov:1971,Haensel:1983,Lindblom:1998dp}
that there is a simple criterion, in terms of the
discontinuity in the energy density at the transition,
that specifies when hybrid stars with an arbitrarily small core (i.e.~with
central pressure just above the transition) will be stable.
In this paper we look at the properties of the hybrid star branch from the
transition up to higher central pressures, assuming that
the quark matter EoS has
a density-independent speed of sound like a classical 
ideal gas\footnote{For a classical ideal gas, the squared speed of sound $c_s^2 \propto T/m$, where $T$ is the temperature and $m$ is the mass, and is independent of density. For a quantum ideal gas, 
$c_s^2 = (1/3)~(1- (m/\mu)^2)$, where $\mu$ is the chemical potential inclusive of mass, and varies between 1/3 (ultra-relativistic case) and $\sim (2/3)~n^{2/3}$ (the non-relativistic case), where $n$ is the number density.
}.
We obtain the phase diagram of the possible topological forms of the
hybrid star branches, and find that it is fairly insensitive to details
of the nuclear matter equation of state. We also investigate the observability
of the hybrid star branches, and
the maximum mass as a function of the parameters of the quark matter EoS.

\begin{figure}[htb]
\includegraphics[width=\hsize]{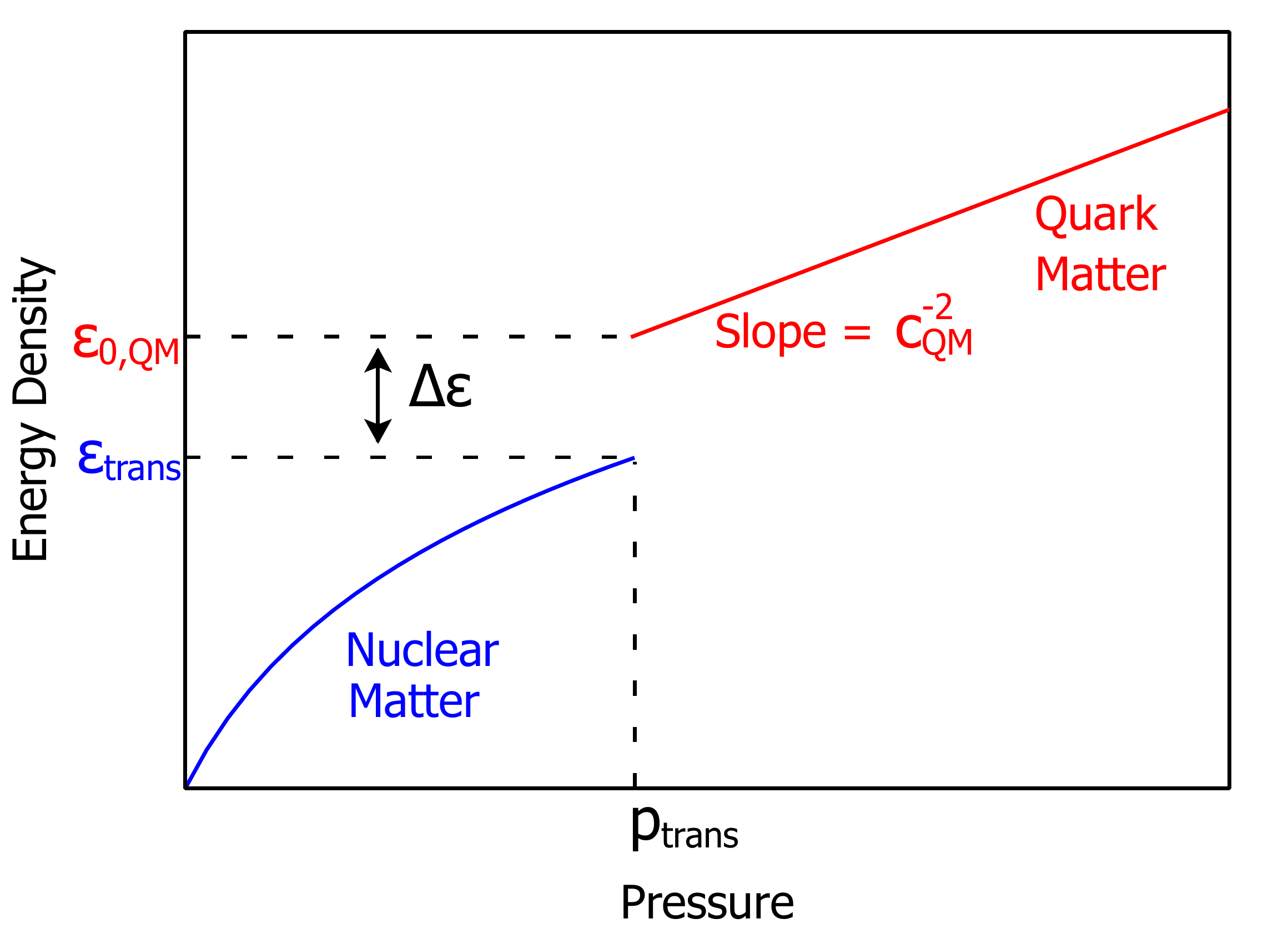}
\caption{Equation of state $\ep(p)$
for dense matter. The quark matter EoS is specified by 
the transition pressure  $\ptrans$, the
energy density discontinuity
$\De\ep$, and the speed of sound in quark matter $\protect\cQM$
(assumed density-independent). 
}
\label{fig:eoshyb}
\end{figure}

To address these questions we
parameterize the quark matter EoS in terms of three
quantities: the pressure $\ptrans$ of the transition from nuclear
matter to quark matter, 
the discontinuity in energy density $\De\ep$ at the transition,
and  the speed of sound $\cQM$ in the quark matter,
which we  assume remains constant as the pressure 
varies from $\ptrans$ up to the central pressure of the
maximum mass star. This ``CSS'' parameterization can be
viewed as the lowest-order terms of a Taylor expansion of the
quark matter EoS about the transition pressure
(see  Fig.~\ref{fig:eoshyb}),
\beq
\ep(p) = \left\{\!
\begin{array}{ll}
\ep_{\rm NM}(p) & p<\ptrans \\
\ep_{\rm NM}(\ptrans)+\De\ep+c_{\rm QM}^{-2} (p-\ptrans) & p>\ptrans
\end{array}
\right.\ ,
\label{eqn:EoSqm1}
\eeq
where $\ep_{\rm NM}(p)$ is the nuclear matter equation of state.
In Appendix~\ref{sec:ideal} we describe the thermodynamically
consistent parameterization of the EoS that we used for
quark matter. A similar generic parameterization was proposed
in Ref.~\cite{Zdunik:2012dj}, which also
considered the possibility of two first-order transitions
involving two different phases of quark matter. See also
Ref.~\cite{Chamel:2012ea}, which set the speed of sound to 1.

\noindent $\bullet$ {\em Quark Matter EoS}.

The assumption that quark matter has
a density-independent speed of sound is
reasonably consistent with some well-known quark matter equations of
state. For some Nambu--Jona-Lasinio models, the CSS EoS fits Eq.~\eqn{eqn:EoSqm1}
almost exactly \cite{Zdunik:2012dj,Agrawal:2010er,Bonanno:2011ch,Lastowiecki:2011hh}.  In addition, the perturbative quark
matter EoS \cite{Kurkela:2010yk} has roughly density-independent $\cQMsq$,
with a value around 0.2 to 0.3, above the transition from nuclear matter (see
Fig.~9 of Ref.~\cite{Kurkela:2009gj}).  In the quartic polynomial
parameterization \cite{Alford:2004pf}, varying the coefficient $a_{2}$ between
$\pm(150 {\rm MeV})^{2}$, and the coefficient $a_{4}$ between 0.6 and 1, and
keeping $\ntrans/n_0$ above 1.5 {($n_0\equiv 0.16\,\fm^{-3}$ is the
nuclear saturation density)}, one finds that $\cQMsq$ is always between 0.3
and 0.36.

In this paper we study hybrid stars
for a range of values of $\cQMsq$, from
1/3 (characteristic of very weakly interacting massless quarks) to 1 (the
maximum value consistent with causality). We expect that this will give us
a reasonable idea
of the likely range of outcomes for realistic quark matter.

\begin{table}

\noindent
\begin{tabular}{l@{\quad}c@{\qquad}cc}
\hline
EoS & max mass & radius at $M=1.4\,M_\odot$ & $L$ \\
\hline
NL3 &  $2.77\,M_\odot$ & 14.92\,km & 118\,MeV \\
HLPS & $2.15\,M_\odot$ & 10.88\,km & ~33\,MeV \\
\hline
\end{tabular}
\label{tab:EoSproperties}
\caption{Properties of the NL3 and HLPS equations of state.
$L$ characterizes the density-dependence of the symmetry energy (see text).
NL3 is an example of a stiff EoS, HLPS is an example of a softer one
at density $n\lesssim 4 n_0$.
}

\end{table}

\begin{figure*}[htb]
\parbox{0.25\hsize}{
{\large ``Absent''}\\
\includegraphics[width=\hsize]{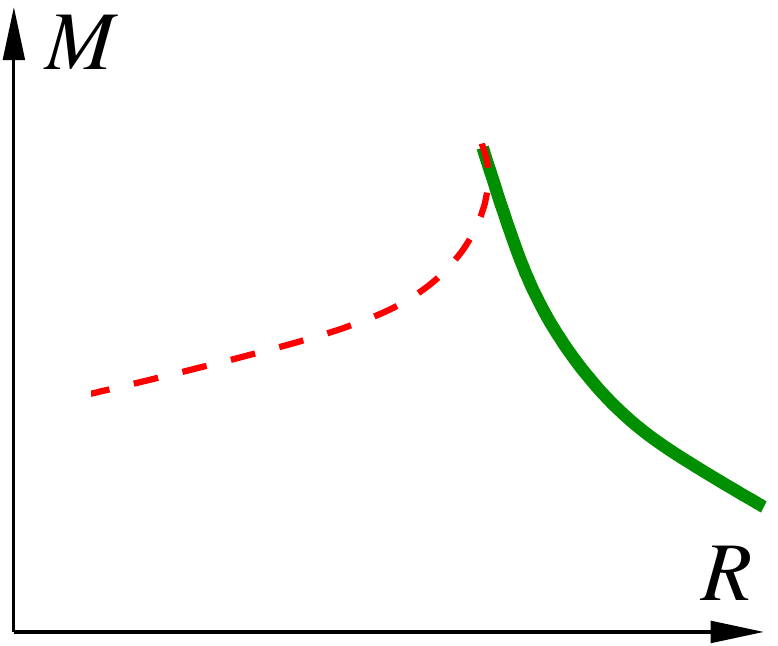}
\bc (a) \ec
}\parbox{0.25\hsize}{
{\large ``Both''}\\
\includegraphics[width=\hsize]{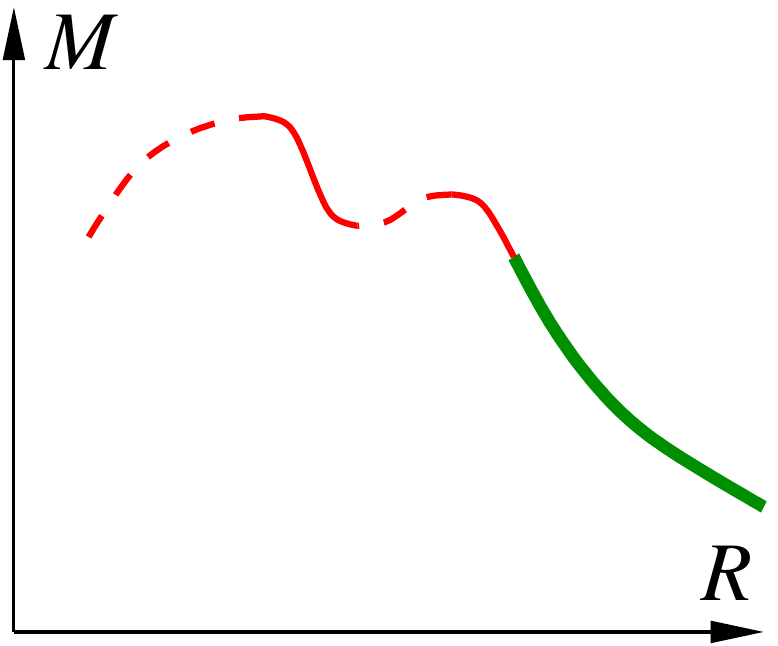}
\bc (b) \ec
}\parbox{0.25\hsize}{
\centerline{\large ``Connected''}
\includegraphics[width=\hsize]{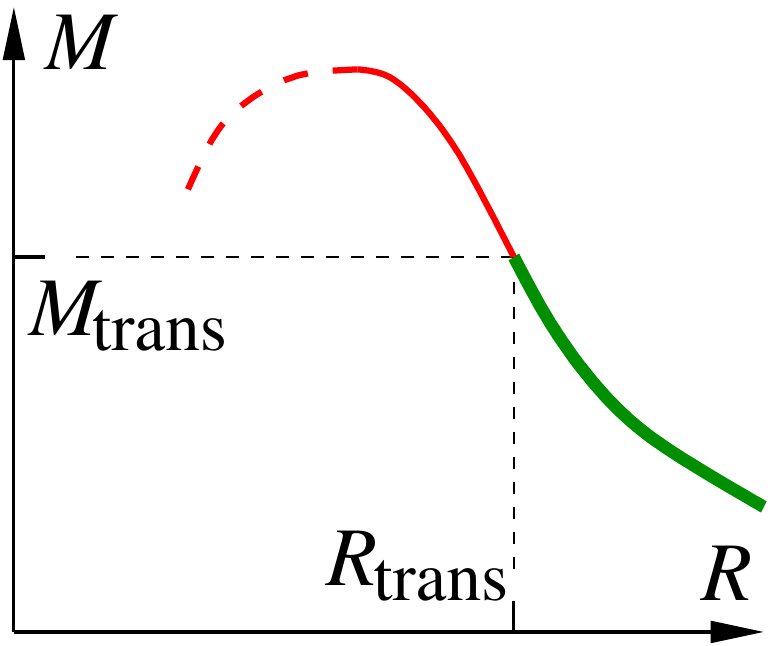}
\bc (c) \ec
}\parbox{0.25\hsize}{
\centerline{\large ``Disconnected''}
\includegraphics[width=\hsize]{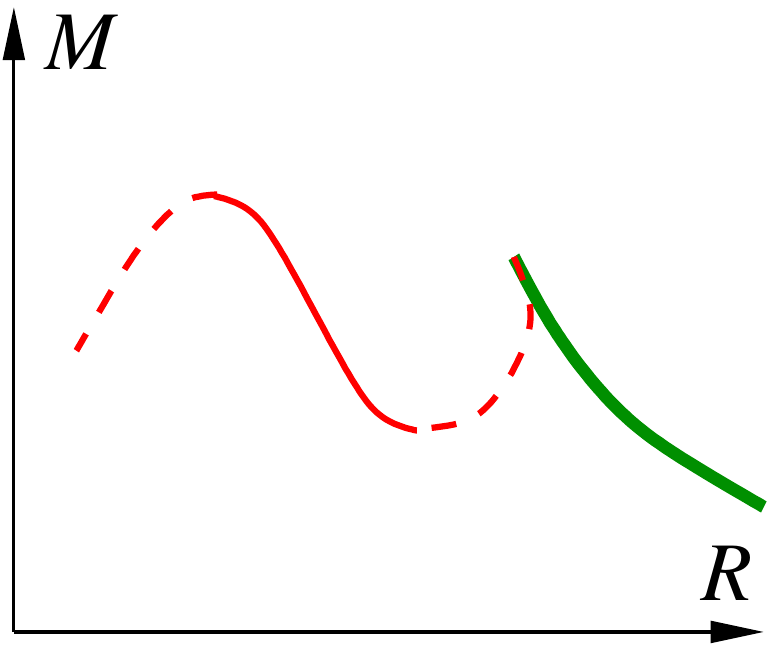}
\bc (d) \ec
}
\caption{
Four possible topologies of
the mass-radius relation for hybrid stars.
The thick (green) line is the hadronic branch. Thin solid (red) lines
are stable hybrid stars; thin dashed (red) lines are unstable hybrid stars.
In (a) the hybrid branch is absent. In (c) there is a connected branch.
In (d) there is a disconnected branch. In (b) there are both types of branch.
In realistic neutron star $M(R)$ curves,
the cusp that occurs in cases (a) and (d) is much smaller and
harder to see \cite{Haensel:1983,Lindblom:1998dp}
}
\label{fig:MR-De}
\end{figure*}

\noindent $\bullet$ {\em Nuclear Matter EoS}.
Up to densities around
nuclear saturation density $n_0$, the nuclear matter EoS can be experimentally
constrained. If one wants to extrapolate it to densities above $n_0$, there
are many proposals in the literature.  For illustrative purposes, we use two
examples: a relativistic mean field model, labelled NL3, \cite{Shen:2011kr} and
a non-relativistic potential model with phenomenological extrapolation
to high density, corresponding to ``EoS1'' in
Ref.~\cite{Hebeler:2010jx}, labelled HLPS.  
Since HLPS is only defined at $n>0.5n_0$, we
continued it to lower density by switching to NL3 for $n<0.5 n_0$.
Some of the properties of HLPS and NL3 are summarized in
Table~\ref{tab:EoSproperties}, where $L$ is related to the derivative of the
symmetry energy $S_2$ with respect to density at the nuclear saturation
density, $L= 3n_0(\partial S_2/\partial n)|_{n_0}$.

HLPS is a softer equation of state, with a lower value of $L$ and lower
pressure at a given energy density (up to $p\approx 3\ee{9}\,\MeV^4$,
$n\approx 5.5\,n_0$ where 
its speed of sound rises above 1 and becomes unphysical).
NL3 is a stiffer EoS, with higher pressure at a given energy density
(also, its speed of sound is less than 1 at all pressures). 
It yields neutron stars that are
larger, and can reach a higher maximum mass.

There is some evidence favoring a soft EoS for nuclear matter:
in Ref.~\cite{Lattimer:2012}, 
values in the range $L=40$ to 60 MeV were favored from an 
analysis of constraints imposed by available laboratory and neutron star data. Using data from X-ray bursts, Ref.~\cite{Steiner:2010} finds 
the surface to volume symmetry energy ratio
$S_s/S_v\approx1.5\pm0.3$ (See after their Eq.~(43) and Table~4), which corresponds to $L$ in the range $22\pm4$ MeV (using Eq.~(7) in Ref.~\cite{Lattimer:2012}).

\noindent $\bullet$ {\em Nuclear/Quark transition}.
The nuclear matter to quark matter transition occurs at
pressure $\ptrans$. We will sometimes specify its position
in terms of the energy density $\etrans$ of nuclear matter at the transition,
or the ratio  $\ptrans/\etrans$.
Since the nuclear matter EoS has $d^2\ep/dp^2<0$ at
high densities (Fig.~\ref{fig:eoshyb}),
$p/\ep$ increases monotonically with $p$, $\ep$, and $n$, so it is
a proxy for the transition pressure or density.

\noindent $\bullet$ {\em Organization of paper}.
In Sec.~\ref{sec:criterion} we discuss the criteria for stable hybrid stars
to exist, as a function of the nuclear matter equation of state and the parameters of our generic quark matter equation of state. Sec.~\ref{sec:phase-diag}
presents the phase diagram for hybrid star branches as a function of the
parameters of the CSS EoS for quark matter.
In Sec.~\ref{sec:maxmass} we discuss the maximum mass that such hybrid stars
can achieve. Sec.~\ref{sec:conclusions} gives a summary and conclusions.

\section{Criterion for stable hybrid stars}
\label{sec:criterion}

\subsection{Connected hybrid branch}

A compact star will be stable as
long as the mass $M$ of the star is an increasing function of the
central pressure $\pcent$ \cite{Bardeen:1966}.
There will therefore be a stable hybrid star branch in 
the $M(R)$ relation, connected to the neutron star branch,
if the mass of the star continues to increase
with $\pcent$ when the quark matter core first appears, at $\pcent=\ptrans$.
When the quark matter core is sufficiently small, its effect on the
star, and hence
the existence of a connected hybrid star branch, is determined
entirely by
the energy density discontinuity $\De \ep$ at its surface, since
the quark core is not large enough for the slope ($c_{\rm QM}^{-2}$) of $\ep(p)$ to have much influence on the mass and radius of the hybrid star.
This fact was pointed out in Ref.~\cite{Seidov:1971} where the dependence
on $\De \ep$ was expressed
in terms of the parameter $q\equiv 1 +\De\ep/\etrans$. A more detailed
treatment in Ref.~\cite{Haensel:1983,Haensel:1987} 
(see also \cite{Migdal:1979je,Kampfer:1981}) used 
a parameter $\la$ with the same definition, and calculated 
the linear response to a small quark matter core in terms of $\la$. 
Ref.~\cite{Lindblom:1998dp} used the parameter 
$\De\equiv \De\ep/(\etrans+\ptrans)$ and highlighted the occurrence of a
cusp in the $M(R)$ relation (see below).

If $\De\ep$ is small then quark matter has a similar energy density 
to that of nuclear matter, and we expect a connected hybrid branch
that looks roughly like a continuation of the nuclear matter branch.
If $\De\ep$  is too large, then 
the star becomes unstable as soon as the quark matter core appears,
because the pressure of the quark matter is unable to counteract the
additional downward force from the
gravitational attraction that the additional energy in the
core exerts on the rest of the star.
By performing an expansion in powers of the size of the quark
matter core, one can show \cite{Seidov:1971,Haensel:1983,Lindblom:1998dp} that
there is a stable hybrid star branch connected to the neutron star
branch if $\De\ep$ is less than a threshold value $\destab$ given by
\beq
\frac{\destab}{\etrans} = \frac{1}{2} + \frac{3}{2}\frac{\ptrans}{\etrans} \ .
\label{eqn:stability}
\eeq
(This is $\la_{\rm crit}-1$ in the notation of Ref.~\cite{Haensel:1983}.)
As $\De\ep$ approaches the threshold value $\destab$ from below, 
both $dM/d\pcent$
and $dR/d\pcent$ approach zero linearly as $p-\pcent$, with the
result that the slope $dM/dR$ of the mass radius curve is independent
of $\De\ep$. For $\De\ep<\destab$, the hybrid star branch
continues
with the same slope as the neutron star mass-radius relation at
the transition to quark matter. When $\De\ep$ exceeds $\destab$,
$dM/dR$ is unchanged, but flips around so that there is a cusp when
the central pressure reaches $\ptrans$ \cite{Lindblom:1998dp}, at
$(M,R)=(\Mtrans,\Rtrans)$. This is shown in schematic form
in Fig.~\ref{fig:MR-De}, where panels (b) and (c) show possible forms
of $M(R)$ for $\De\ep<\destab$, and panels (a) and (d) show possible forms
for $\De\ep>\destab$. In $M(R)$ curves for realistic neutron star equations
of state, the cusp at high $\De\ep$ is much less clearly visible: 
the region where the
slopes of hybrid and neutron stars match is very small, covering a range
in $M$ of less than one part in a thousand near the transition 
point $(\Rtrans,\Mtrans)$ \cite{Lindblom:1998dp}.

\subsection{Disconnected hybrid branch}

In Figs.~\ref{fig:MR-De}(b) and (d), we illustrate the occurrence of a second,
disconnected, branch of stable hybrid stars at $\De\ep > \destab$. 
This possibility was noted in Ref.~\cite{Haensel:1983}.
The disconnected branch is a ``third family''
\cite{Glendenning:2000,Schertler:2000} of compact stars besides neutron stars
and white dwarfs. Third families have been found in $M(R)$ calculations for
specific quark matter models, for example kaon condensed stars
\cite{Thorsson:94}, quark matter cores from perturbative QCD
\cite{Fraga:2001id}, and color superconducting quark matter cores
\cite{Banik:2003}.  In
this paper we
will study the generic features of a quark matter EoS that give rise to this
phenomenon. In principle one could imagine that additional disconnected
stable branches might occur, but we do not find any with the CSS
parameterization of the quark matter EoS.

\begin{figure*}[htb]
\parbox{0.5\hsize}{
\centerline{\large HLPS + CSS$(c^2\!=\!1)$}
\includegraphics[width=\hsize]{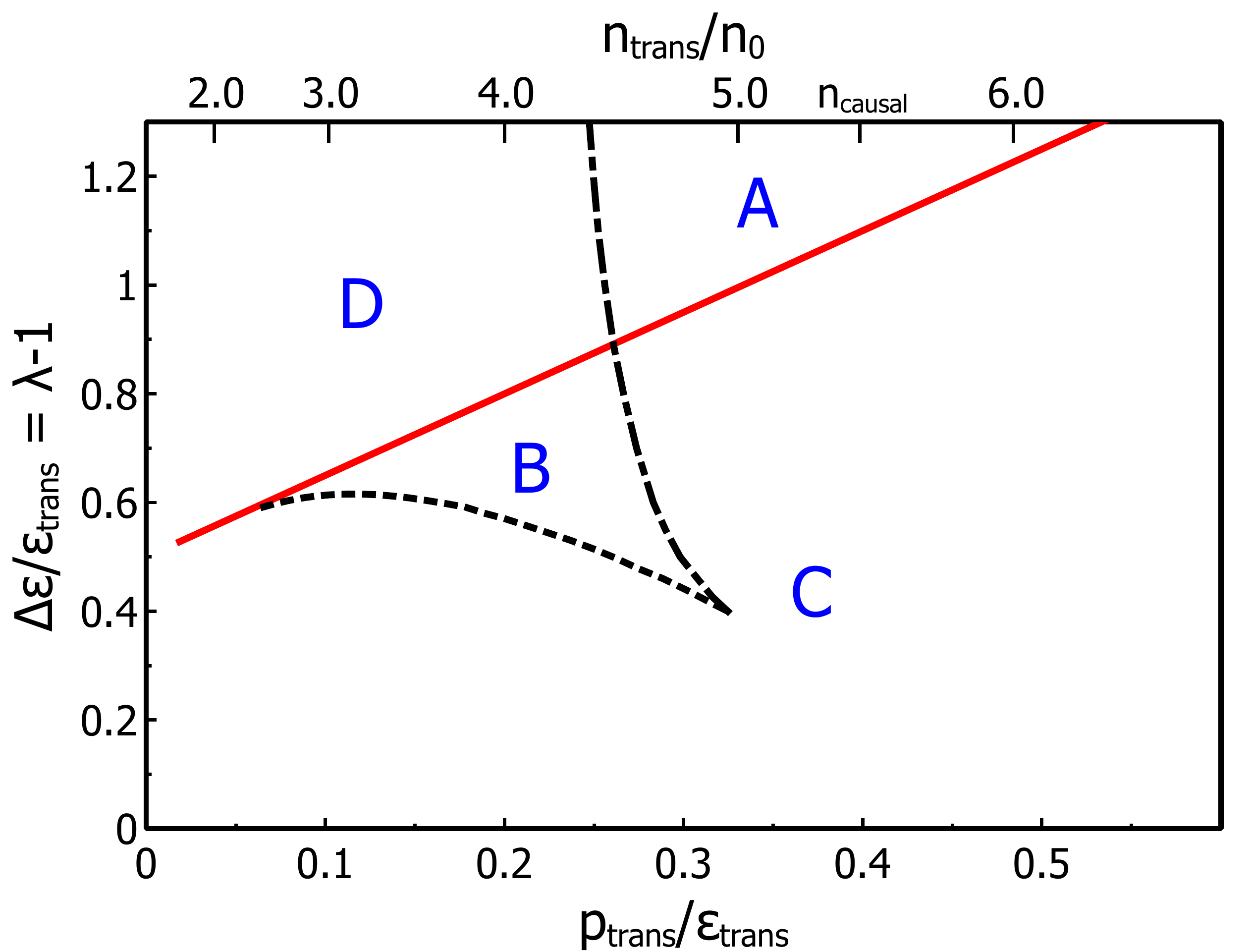}\\[2ex]
\centerline{\large (a)}
}\parbox{0.5\hsize}{
\smallskip
\includegraphics[width=0.9\hsize]{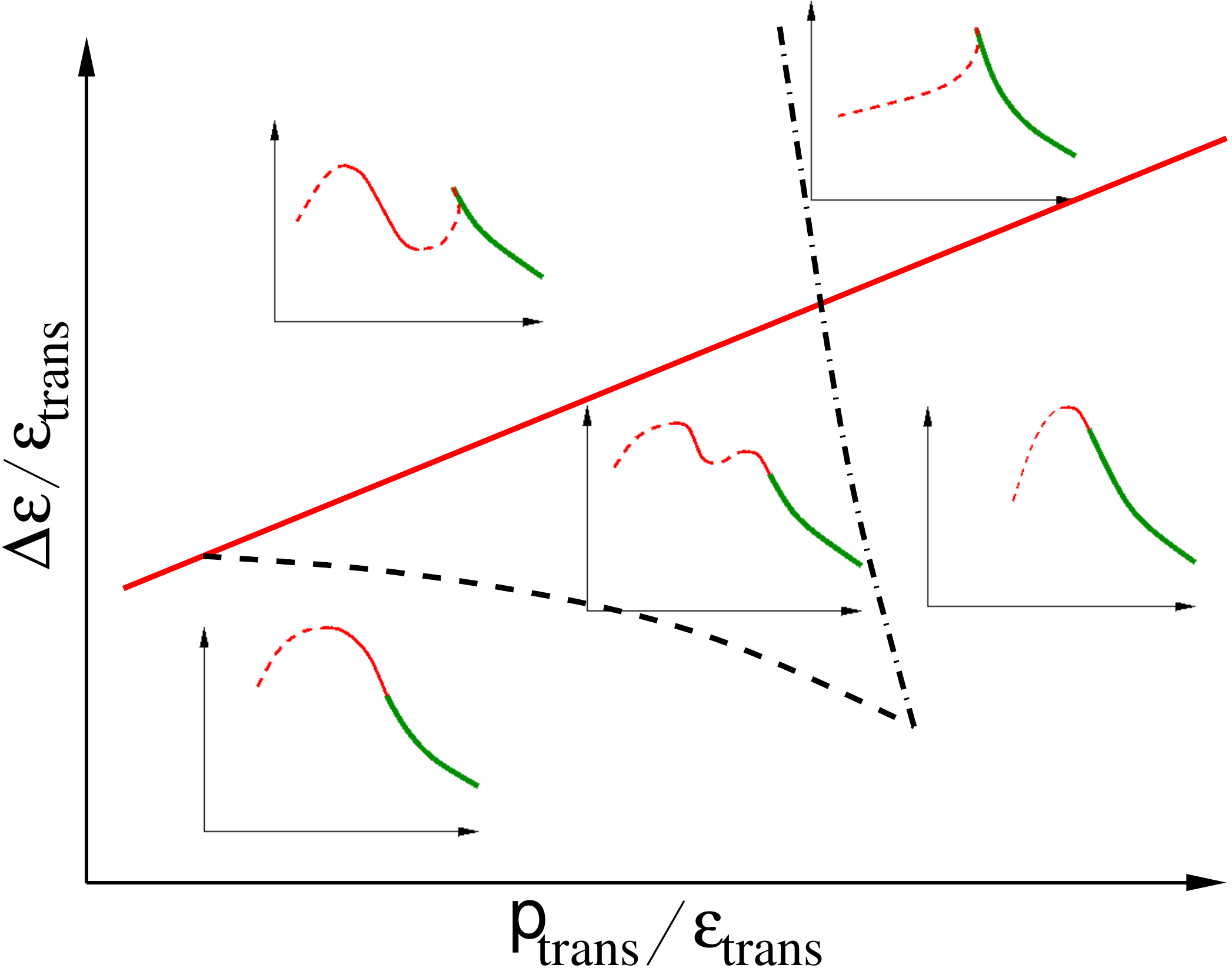}\\[2ex]
\centerline{\large (b)}
}
\caption{Phase diagram for hybrid star branches in the
mass-radius relation of compact stars. The control parameters are the 
pressure $\ptrans$ and energy density discontinuity $\De \ep$ at the transition,
each expressed in units of the nuclear energy density at the transition $\etrans$.
The y-axis is therefore just a shifted version of the parameter $\la$ 
of Ref.~\cite{Haensel:1983}.
The solid straight line is $\destab$ (Eq.~\eqn{eqn:stability}).
The left panel is the result of calculations for a 
softer nuclear matter EoS (HLPS) and CSS quark matter
with $\cQMsq=1$.
The right panel is a schematic showing the topological
form of the mass-radius
relation in each region of the diagram:
regions A,B,C,D correspond to Fig.~\ref{fig:MR-De}(a),(b),(c),(d).
}
\label{fig:phase-diag-HLPS}
\end{figure*}

\begin{figure*}
\parbox{0.5\hsize}{
\centerline{\large HLPS \& NL3 + CSS$(c^2\!=\!1/3)$ }
\vspace{-2ex}
\includegraphics[width=\hsize]{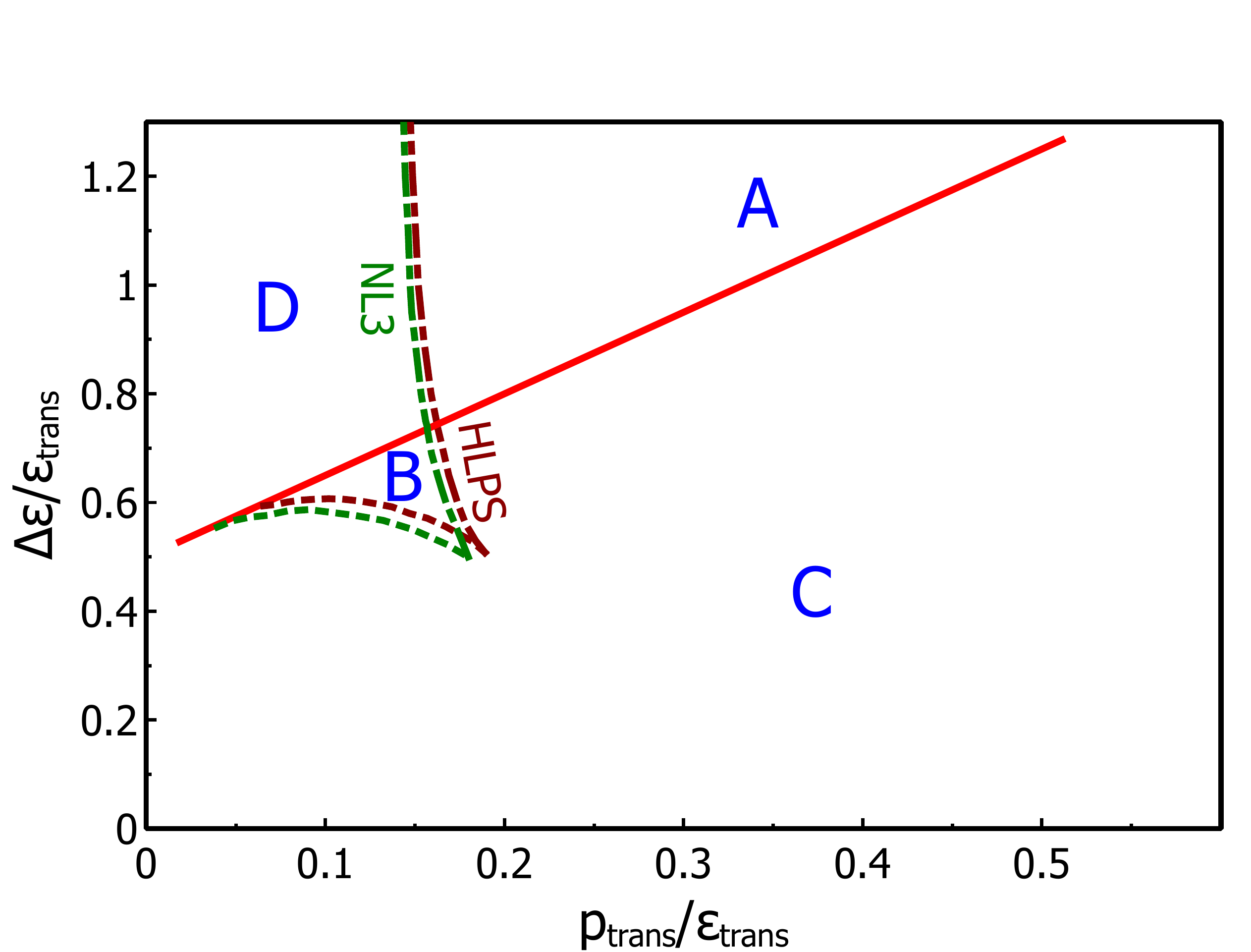}\\[2ex]
\centerline{\large (a)}
}\parbox{0.5\hsize}{
\centerline{\large HLPS \& NL3 + CSS$(c^2\!=\!1)$}
\vspace{-2ex}
\includegraphics[width=\hsize]{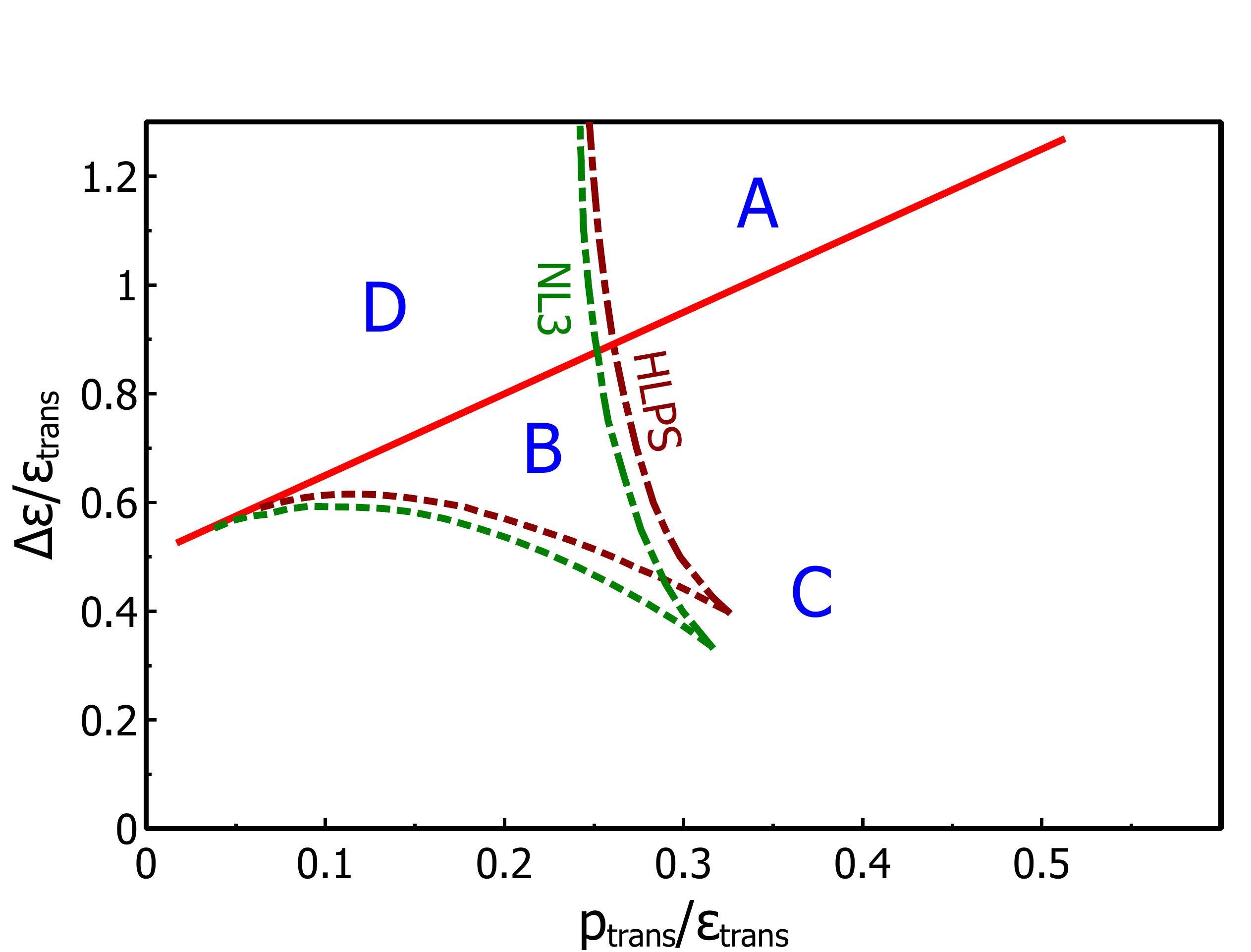}\\[2ex]
\centerline{\large (b)}
}
\caption{Phase diagram like  Fig.~\ref{fig:phase-diag-HLPS}, showing that
the phase boundaries are not very sensitive to changes in the nuclear 
matter EoS, but they are affected by varying the quark matter speed of sound.
}
\label{fig:phase-diag-both}
\end{figure*}

\section{``Phase diagram'' for hybrid stars}
\label{sec:phase-diag}

\subsection{Phase diagram at fixed $\protect\cQM$}

In Fig.~\ref{fig:phase-diag-HLPS} we plot a ``phase diagram'' for hybrid stars, 
where the control parameters are two of the parameters of the
CSS quark matter EoS: $\ptrans/\etrans$
(the ratio of pressure to energy density in nuclear matter
at the transition pressure) and  $\De\ep$
(the discontinuity in the energy density at the
transition). The quark matter speed of sound is held constant.
As noted in Sec.~\ref{sec:intro},
$\ptrans/\etrans$ increases monotonically with pressure, so it is 
a proxy for the transition pressure or density.
Our phase diagram covers a range in $\ptrans/\etrans$ from 0.02 up to about 0.5.
Below 0.02 the NL3 EoS has a baryon density far below $n_0$, and the
HLPS EoS has a discontinuity in $c^2$ that is not physical.

The left panel of Fig.~\ref{fig:phase-diag-HLPS}
is the result of calculations for the 
HLPS nuclear matter EoS and quark matter with $\cQMsq=1$.
The lower x-axis shows $\ptrans/\etrans$, which is the natural way to
characterize the transition. The upper x-axis shows 
the corresponding transition baryon density of nuclear matter
for the HLPS EoS. HLPS becomes acausal at  $n=5.458\,n_0$, but
all the interesting structure in the plot is below this density.
The right panel is a schematic showing the form of the mass-radius
relation in each region of the diagram. The regions correspond to
different topologies of the hybrid branch displayed
in Fig.~\ref{fig:MR-De}:
A=``Absent'', C=``Connected'', D=``Disconnected'', B=``Both''
(connected and disconnected).

The solid straight (red) line is $\destab$ from Eq.~\eqn{eqn:stability}.
On crossing this line $M'(\pcent)$ and $R'(\pcent)$ both change sign
when quark matter first appears, causing the $M(R)$ curve to undergo
a discontinuous change, flipping around from being ``upward-pointing''
(mass increases with central pressure) and continuous,
to being ``downward-pointing''
(mass decreases with central pressure) with a cusp. 
Below the line ($\De\ep<\destab$), in regions B and C, there is a hybrid star
branch connected to the nuclear star branch. Above the line ($\De\ep>\destab$),
in regions A and D, there is no connected
hybrid star branch. In regions B and D there is
a disconnected hybrid star branch.

The roughly vertical dash-dotted curve in  Fig.~\ref{fig:phase-diag-HLPS}
marks a transition where an additional disconnected branch of hybrid
stars appears/disappears. When one crosses this line from the right,
going from region A to D by decreasing the nuclear/quark
transition density, a stationary point of inflection appears in $M(\pcent)$
at $\pcent>\ptrans$.
If one crosses from C to B then this point of inflection is at lower
central pressure than the existing maximum in $M(\pcent)$. This 
produces a stationary point of inflection 
in the $M(R)$ relation to the left of the existing maximum (if any).
After crossing the dash-dotted line the point of inflection
becomes a new maximum-minimum pair (the maximum being further from the
transition point), producing a disconnected branch of stable hybrid stars
in regions B and D.
Crossing the other way, by increasing the transition pressure,
the maximum and minimum that delimit the
disconnected branch merge and the branch disappears.

The roughly horizontal dashed curve
in Fig.~\ref{fig:phase-diag-HLPS} which separates region B and C 
marks a transition between mass-radius relations with one connected
hybrid star branch, and those with two hybrid star branches, one connected
and one disconnected. In the notation 
of Ref.~\cite{Zdunik:2005kh} (sec. (4.2)), this line corresponds to 
$\De\ep/\etrans = \la_{\rm max}-1$.
Crossing this line from below, by increasing
the energy density discontinuity, 
a stationary point of inflection in $M(\pcent)$
(or equivalently in $M(R)$) appears in the existing
connected hybrid branch.
Crossing in to region B, this point of inflection
becomes a new maximum-minimum pair, so the connected hybrid branch
is broken in to a smaller connected branch and a new disconnected branch.
The maximum of the old connected branch smoothly becomes the maximum
of the new disconnected branch.
If one crossed the dashed line in the 
opposite direction, from B to C, the
maximum closest to the transition point would approach the minimum
and they would annihilate, leaving only the more distant maximum.

Where the horizontal and vertical curves meet, the two maxima and the minimum
that are present in region B all merge to form a single flat maximum
where the first three derivatives of $M(R)$ are all zero.

Along the critical line Eq.~\eqn{eqn:stability},
which is the  straight line in Fig.~\ref{fig:phase-diag-HLPS}, 
$dM/d\pcent=0$ at
$\pcent=\ptrans$. We now discuss the curvature 
$M''\equiv d^2M/d\pcent^2$ at $\pcent=\ptrans$ 
on that line.
On the A/C boundary, $M''<0$. In region C
there is a maximum in $M(\pcent)$ at $\pcent>\ptrans$, 
but on the A/C boundary that maximum has shifted to $\pcent=\ptrans$.
As we move down the A/C boundary, $M''$ becomes less negative. This
continues along the B/D boundary, until at the point
where the B/C boundary (dashed line) merges with the B/D boundary, $M''=0$.
This is because the stationary point of inflection in $M(\pcent)$ 
(where the disconnected branch first appears)
has now arrived at $\pcent=\ptrans$. Continuing down
the critical line, $M''$ becomes positive, making it possible for
there to be a direct transition between C and D.

\subsection{Varying $\protect\cQM$ and the nuclear EoS}

Up to now we have discussed the effects of varying two of the parameters
of the quark matter EoS, namely $\ptrans$ and $\De\ep$.
In Fig.~\ref{fig:phase-diag-both} we show the effects of varying the
third parameter, the speed of sound, and the effect of varying the
nuclear matter EoS.
Fig.~\ref{fig:phase-diag-both}(a) is the phase diagram for 
CSS quark matter with $\cQMsq=1/3$,
and Fig.~\ref{fig:phase-diag-both}(b) is for $\cQMsq=1$. In both panels
we show the phase diagram for 
a softer (HLPS) nuclear matter EoS and a harder (NL3) nuclear matter EoS. 
As expected from Eq.~\eqn{eqn:stability} the 
straight line where the connected hybrid branch
disappears is independent of $\cQMsq$ and the detailed form
of the nuclear matter EoS.
The other phase boundaries, outlining the region where there is a
disconnected hybrid branch, are remarkably insensitive to the
details of the nuclear matter EoS. However, they depend strongly on
the quark matter speed of sound. For a given nuclear matter EoS
the hybrid branch
structure is determined by $\ptrans/\etrans$, $\De\ep/\etrans$, and
$\cQMsq$, so one could make a three-dimensional plot with $\cQMsq$ as
the third axis, but this figure adequately illustrates the dependence 
on $\cQMsq$. We will now discuss the physics behind the shape of the
phase boundaries.

\begin{figure*}[htb]
\parbox{0.5\hsize}{
\centerline{\large HLPS + CSS$(c^2\!=\!1)$}
\includegraphics[width=\hsize]{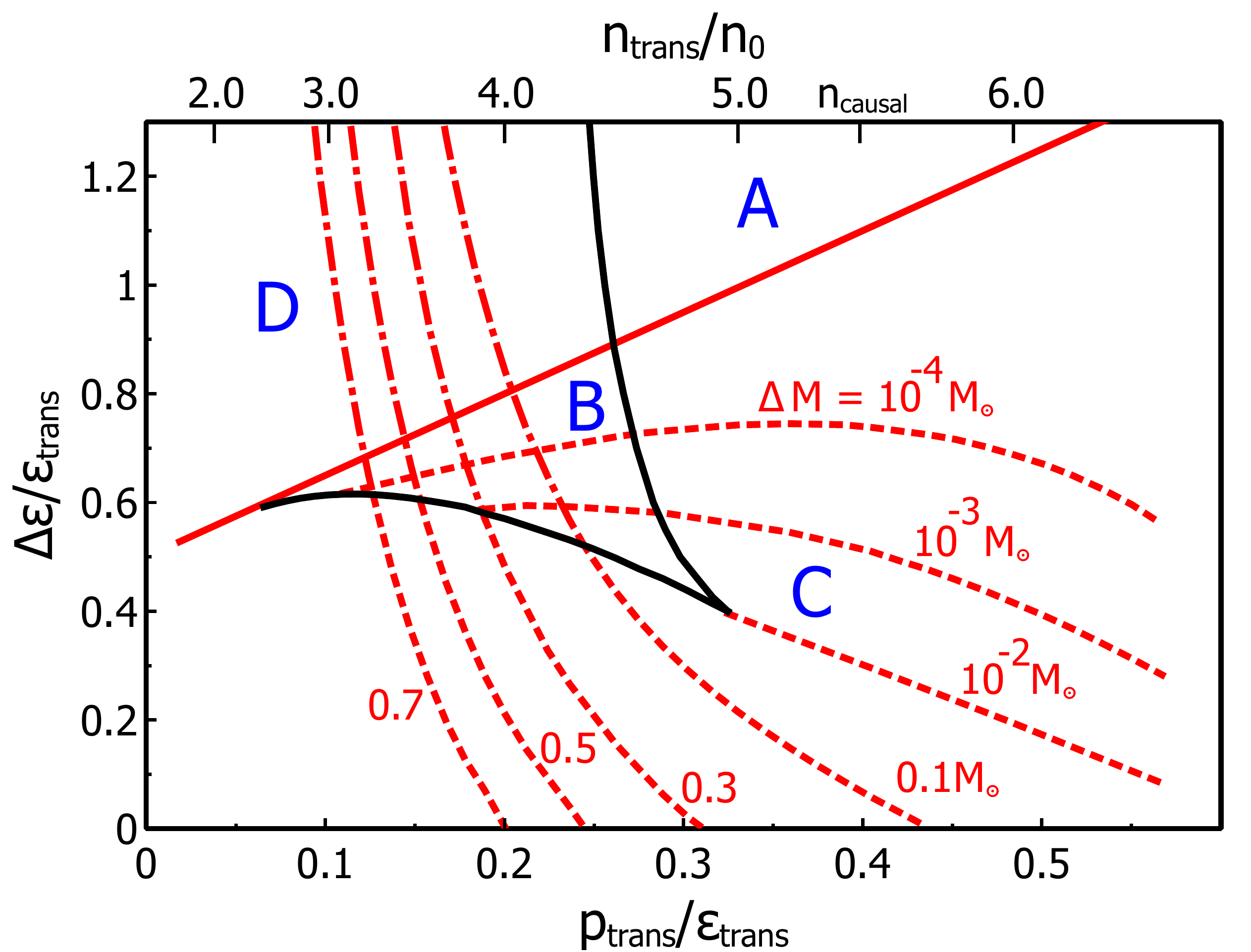}\\[2ex]
\centerline{\large (a)}
}\parbox{0.5\hsize}{
\centerline{\large NL3 + CSS$(c^2\!=\!1)$}
\includegraphics[width=\hsize]{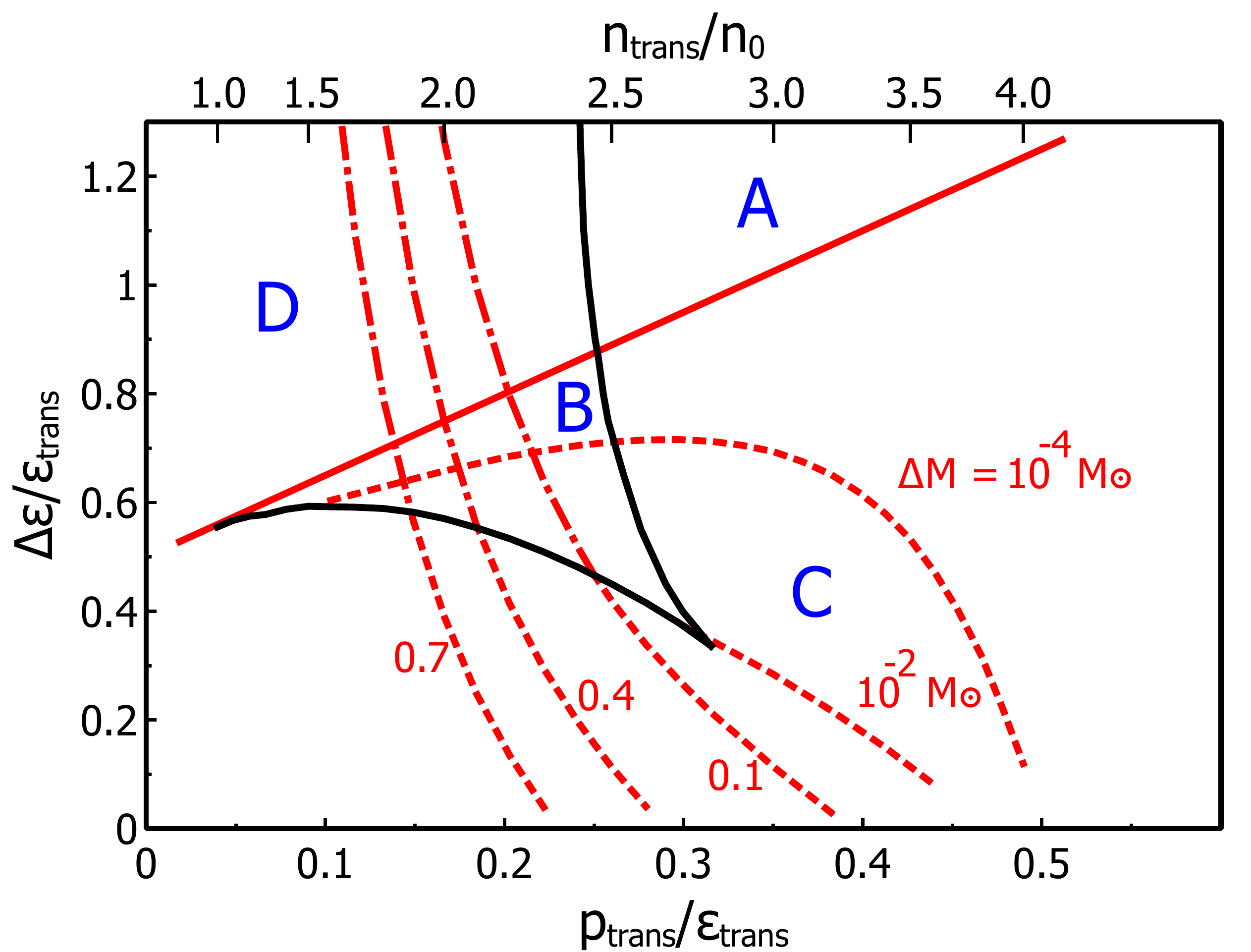}\\[2ex]
\centerline{\large (b)}
}
\caption{Contour plot of a measure of the observability of hybrid branches:
$\De M$, the mass difference between the heaviest hybrid star and the
hadronic star when quark matter first appears.
We show results for HLPS and NL3 nuclear matter, 
with $\cQMsq=1$ CSS quark matter.
The contours are not very sensitive to details of the nuclear matter EoS.
If the transition density is high, or if a disconnected branch is present,
the connected branch may be very small and hard to observe.
}
\label{fig:phase-diag-Mcontours}
\end{figure*}

\begin{figure*}[htb]
\parbox{0.5\hsize}{
\centerline{\large HLPS + CSS$(c^2\!=\!1)$}
\includegraphics[width=\hsize]{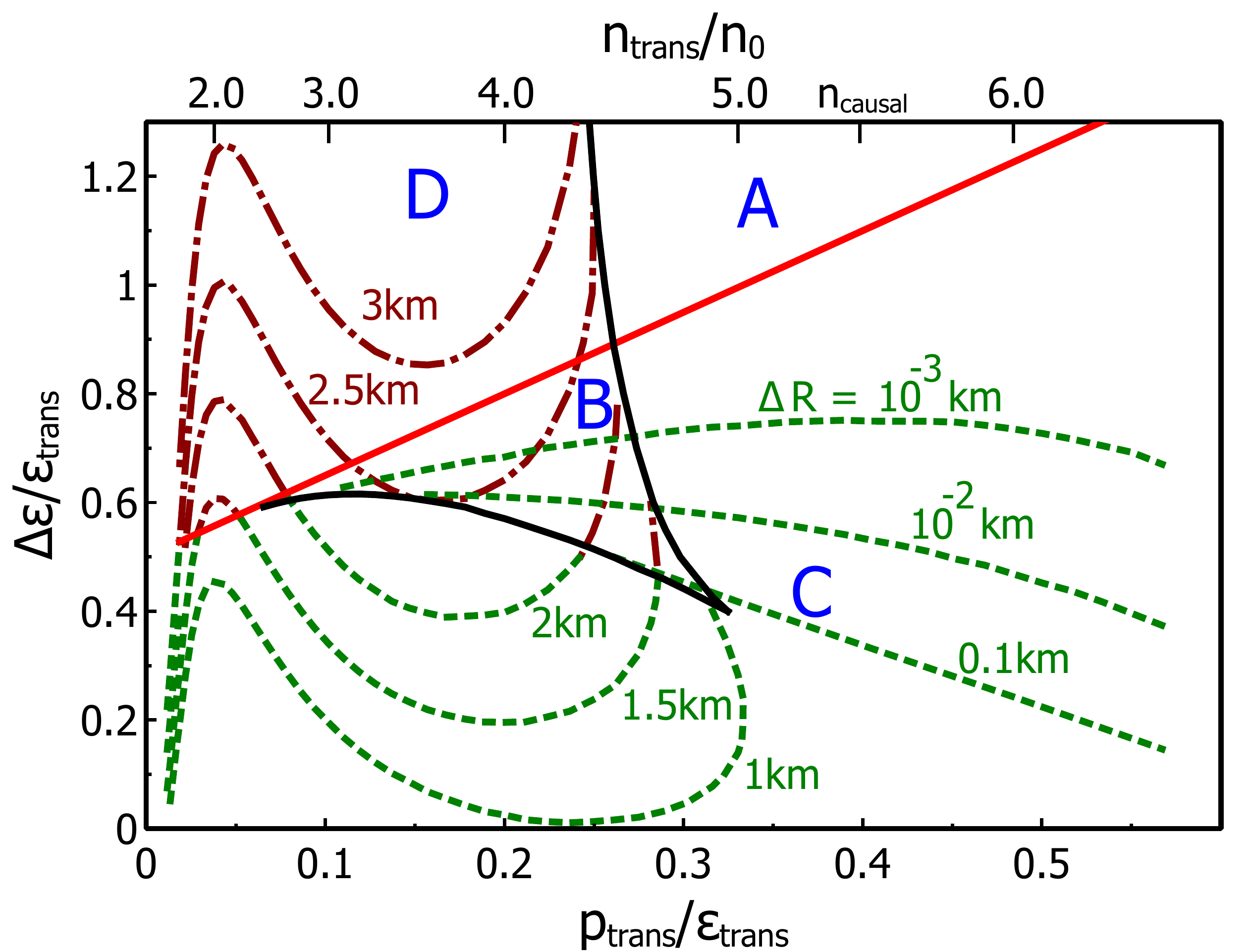}\\[2ex]
\centerline{\large (a)}
}\parbox{0.5\hsize}{
\centerline{\large NL3 + CSS$(c^2\!=\!1)$}
\includegraphics[width=\hsize]{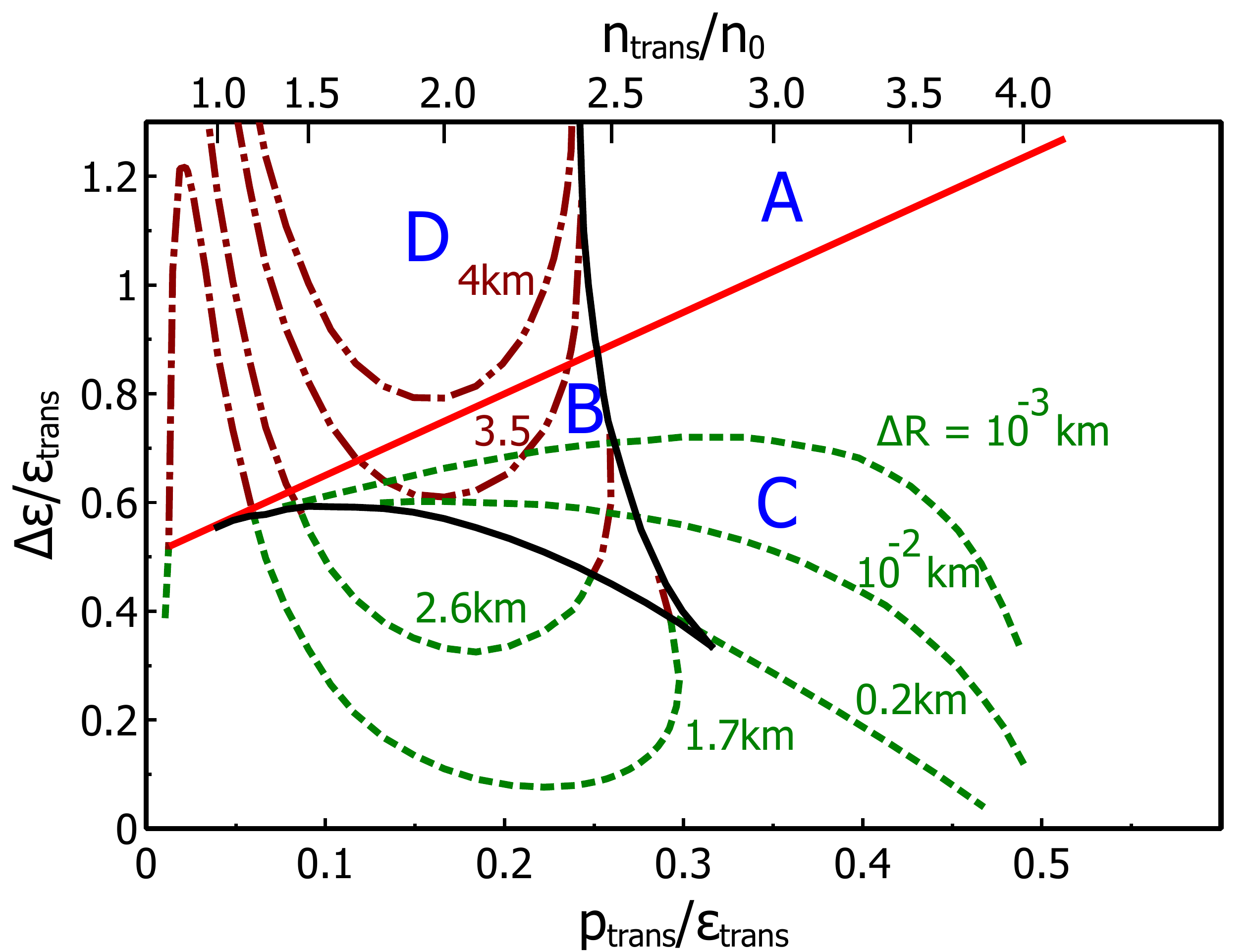}\\[2ex]
\centerline{\large (b)}
}
\caption{Contour plot of a measure of the observability of hybrid branches:
 $\De R$, the difference between the radius
of the hadronic star when quark
matter first appears and the radius of the
heaviest hybrid star. 
We show results for HLPS and NL3 nuclear matter,
with $\cQMsq=1$ CSS quark matter.
The contours are not very sensitive to details of the nuclear matter EoS.
If the transition density is high, or if a disconnected branch
is present, the connected branch may be very small
and hard to observe.
}
\label{fig:phase-diag-Rcontours}
\end{figure*}

\subsection{Physical understanding of the phase diagram}

The main feature of the phase diagram is that
a disconnected branch is present when
the transition density is sufficiently low, and the energy density discontinuity
is sufficiently high, namely in regions B and D.
It occurs more readily (i.e.\ those regions are larger)
if the speed of sound in quark matter is high.
Such features were noticed in the context of stars with mixed phases
in Ref.~\cite{Macher:2004vw}, which pointed out that they
can be understood as follows.

When a very small quark matter core is present, its greater density creates
an additional gravitational pull on the nuclear mantle. If the pressure of the
core can counteract the extra pull, the star is stable
and there is a connected hybrid branch (regions C and B).
If the energy density jump is too great, the extra gravitational pull is too
strong, and the star becomes unstable when quark matter first appears
(regions A and D). However, if
the energy density of the core rises slowly enough with increasing pressure
(i.e. if  $c^2=dp/d\ep$ is large enough), a larger core with
a higher central pressure may be able to sustain the weight of the nuclear
mantle above it (region D). Region B, with connected and disconnected branches,
is more complicated and we do not have an intuitive explanation for it.


We can now understand why the
vertical line marking the B/C and D/A boundaries moves to the right as $\cQMsq$
increases. Since $c^2=dp/d\ep$, if $\cQMsq$
is larger then the energy density of the core rises 
more slowly with increasing pressure, which minimizes the tendency for
a large core to destabilize the star via its gravitational attraction.
Finally, we can see why that line has a slight negative slope: larger
$\De\ep$ makes the quark core heavier, increasing its pull on the nuclear 
mantle, and making the hybrid star more unstable against collapse.

Ref.~\cite{Macher:2004vw}, assuming a mixed phase, conjectured that
the third family (disconnected branch) 
exists when the speed of sound in quark matter
is higher than that in the  mixed phase. However, our results do not
support that conjecture. In the case we study,
with no mixed phase, the relevant quantity would be the difference
between $\cQM$ (the speed of sound in quark matter) and $\cNM$
(the speed of sound
in nuclear matter at the phase transition).
If the conjecture were correct,
our phase diagram (which is for fixed $\cQM$)
would show the disconnected branch appearing at a vertical phase boundary located
at the transition pressure where $\cNM=\cQM$.
In fact, as seen in Fig.~\ref{fig:phase-diag-both},
we find one horizontal phase boundary, and a near-vertical boundary.
For  $\cQMsq=1$ quark matter the vertical boundary occurs at
$\ptrans/\etrans\approx 0.3$ where $\cNMsq\approx0.75$ for HLPS and
$\cNMsq\approx0.65$ for NL3.
For  $\cQMsq=1/3$ quark matter the vertical boundary occurs at
$\ptrans/\etrans\approx 0.15$ where $\cNMsq\approx0.5$ for HLPS and NL3.
We conclude that the appearance of a disconnected branch is not
determined by whether $\cQM>\cNM$.

\begin{figure}
\includegraphics[width=\hsize]{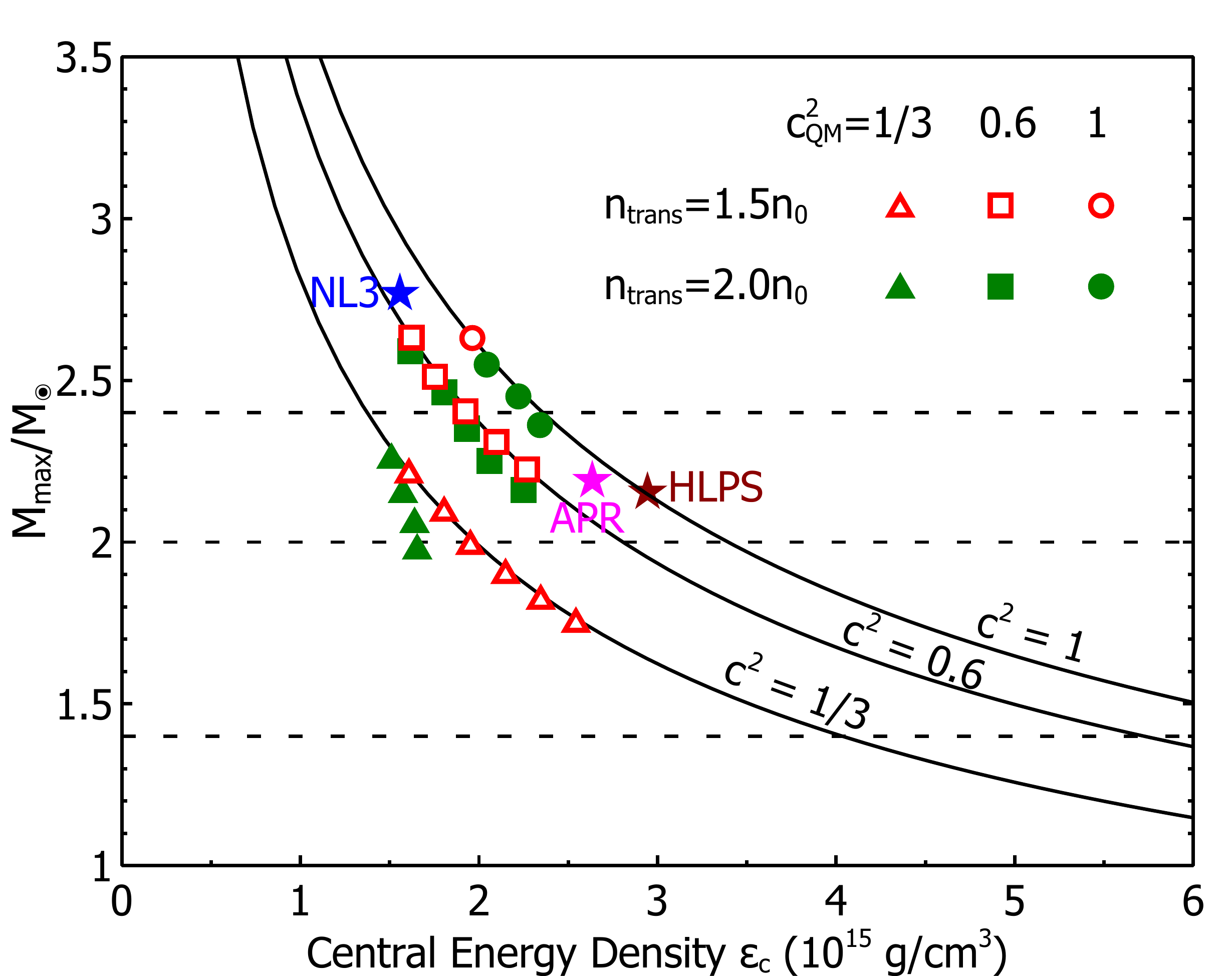}
\caption{Mass of the heaviest hybrid star as a function of its central 
energy density, for various quark matter equations of state \eqn{eqn:EoSqm1}.
The curves are 
predictions of Ref.~\cite{Lattimer:2010uk} for 
stars whose core-region speed of sound squared is $1$, $0.6$, and $1/3$.
Pure nuclear matter stars for the NL3 and APR equations of state
are also plotted. 
}
\label{fig:Mm-ec}
\end{figure}

\subsection{Observability of hybrid branches}

The phase diagrams of the previous subsection show us when
connected and
disconnected branches are present, but for astrophysical observations
it is important to know how easily these branches could be detected
via measurements of the mass and radius of compact stars.
In Figs.~\ref{fig:phase-diag-Mcontours} and \ref{fig:phase-diag-Rcontours}
we show the phase diagram for HLPS and NL3 nuclear matter, 
with $\cQMsq=1$ quark matter, with contours
showing two measures, $\De M$ and $\De R$, of the length of the hybrid branch.
$\De M$ is the difference in mass between the heaviest hybrid star and
the hadronic star just before quark matter appears (whose mass is $\Mtrans$).
$\De R$ is the difference in radius between
the hadronic star just before quark matter appears (whose radius is $\Rtrans$)
and the heaviest hybrid star.
At the horizontal boundary between region C and B, the 
maximum of the connected branch smoothly becomes the maximum of the disconnected
branch (see Fig.~\ref{fig:phase-diag-HLPS}) so the dashed contours 
(for the connected branch) connect smoothly
to the dot-dashed contours (for the disconnected branch).
We see that the $\De M$ and $\De R$ contours are roughly independent
of details of the nuclear matter EoS, except at high transition pressure
where the transition to quark matter is happening close to the maximum of the
nuclear EoS, which greatly suppresses the length of the connected branch.
Note that although we only show contours for positive  $\De M$ and $\De R$,
contours that end on the A/D or near-vertical B/C boundary can 
have negative $\De M$.

From Figs.~\ref{fig:phase-diag-Mcontours} and \ref{fig:phase-diag-Rcontours}
we conclude that Eq.~\eqn{eqn:stability} is not
a good guide to the presence of {\it observable} hybrid branches.
The connected branch may be very small and hard to
detect, even at values of
$\De\ep/\etrans$ well below the critical value of  Eq.~\eqn{eqn:stability}.
For the $\cQMsq=1$ case,
only in the parts of region C that lie approximately below region B
(i.e. for  $\ptrans/\etrans\lesssim 0.3$) is the connected
hybrid branch large enough to be detectable via observations of mass that have an experimental uncertainty of around $0.01\Msolar$ or observations of radius
that have an uncertainty of around $0.2\,\km$.
When a disconnected branch is present,
the connected branch is either absent (region D) or too small to observe
(region B). 

The disconnected branch itself, whose presence has nothing to do with
Eq.~\eqn{eqn:stability}, is in principle easily observable except perhaps at
the far right edge of regions B and D. The natural way for the disconnected
branch to be populated would be via accretion, taking a star to the top of the
connected branch, after which it would have to collapse to the disconnected
branch, with dramatic emission of neutrinos and gamma rays
\cite{Mishustin:2002xe} and gravitational waves \cite{Lin:2005zda}.  It is not
clear how one would populate the parts of the disconnected branch that lie
below that mass threshold, so one might end up with a gap in the observed
radii of neutron stars, where the populated sequence jumps from one branch to
the other.

\begin{figure*}[htb]
\parbox{0.5\hsize}{
\centerline{\large HLPS + CSS}
\includegraphics[width=\hsize]{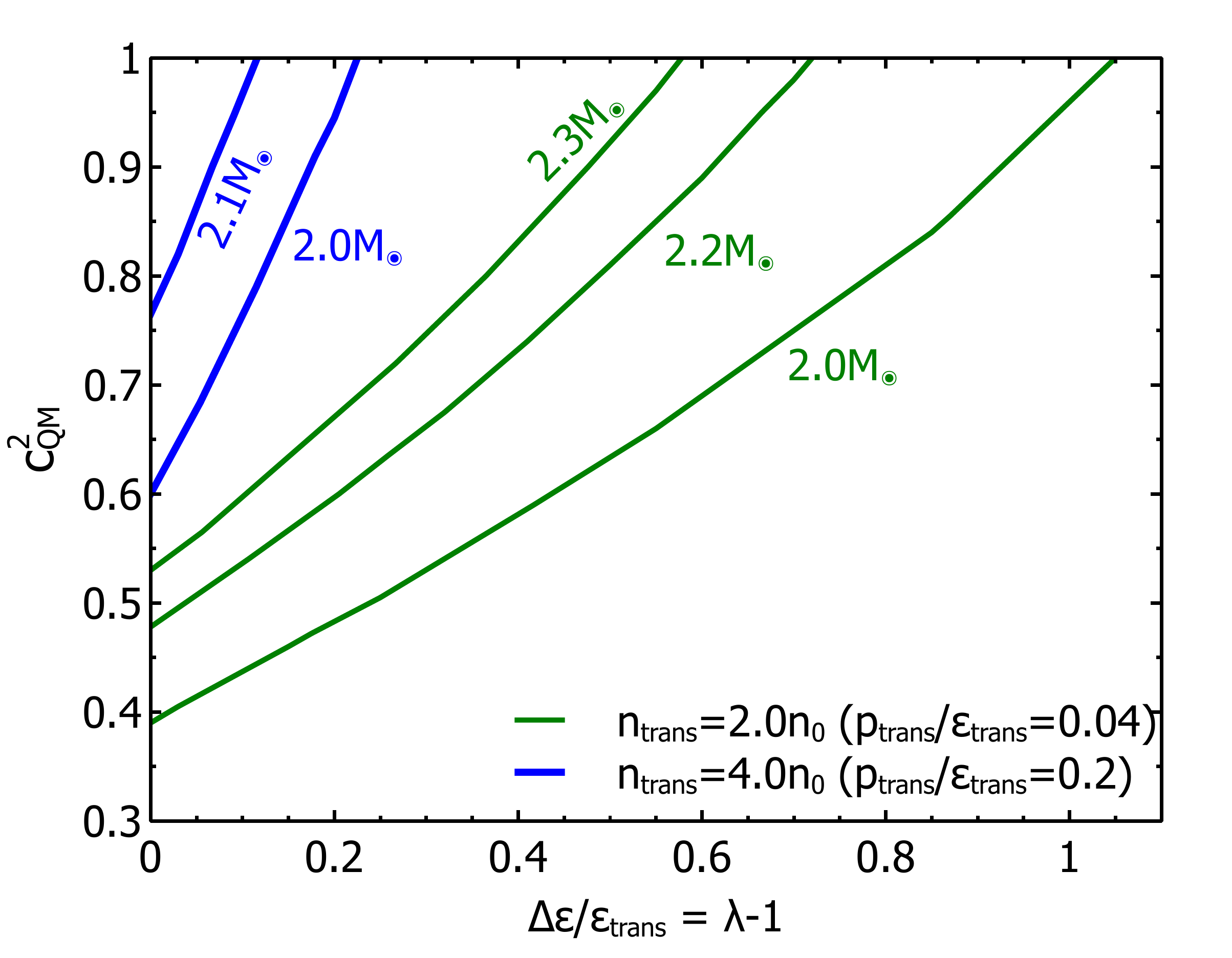}
\centerline{\large (a)}
}\parbox{0.5\hsize}{
\centerline{\large NL3 + CSS}
\includegraphics[width=\hsize]{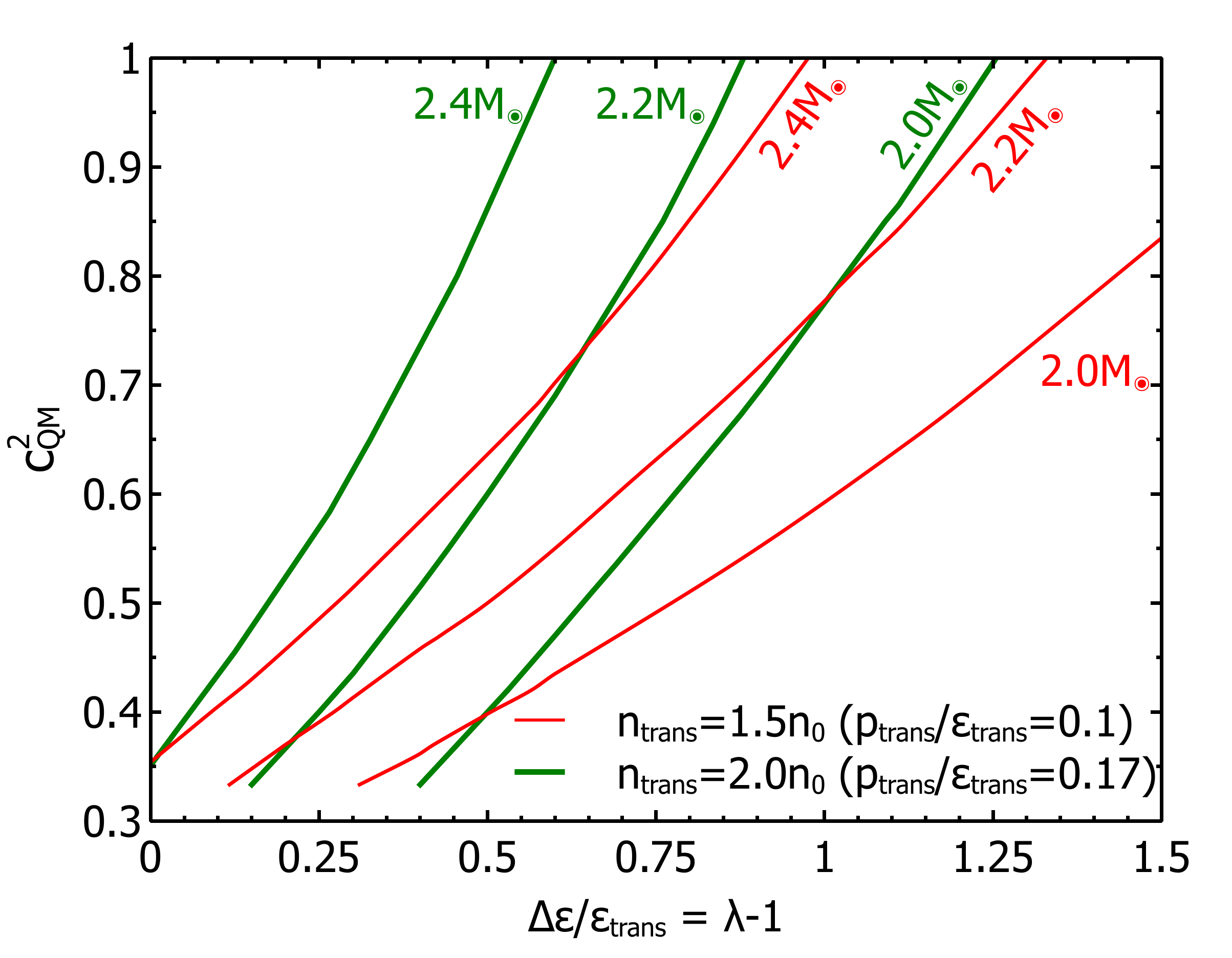}
\centerline{\large (b)}
}
\caption{Contour plot of the mass of the heaviest hybrid star as a function of 
quark matter EoS parameters $\ptrans/\etrans$, $\cQMsq$, and $\De\ep/\etrans$
(a shifted version of $\la$ in Ref.~\cite{Haensel:1983})
for HLPS and NL3 nuclear matter.
The thin (red), medium (green) and thick (blue) lines are for a nuclear to
quark transition at $\ntrans=1.5 n_0$, $2 n_0$, and $4n_0$, respectively.
}
\label{fig:maxM-QM-params}
\end{figure*}

\section{Maximum mass of hybrid stars}
\label{sec:maxmass}

\subsection{Maximum mass and central energy density}

The CSS quark matter EoS \eqn{eqn:EoSqm1} involves
three parameters, $\ptrans$ (or equivalently $\ntrans$), 
$\De\ep$, and $\cQM$.
The results of Ref.~\cite{Lattimer:2004sa,Lattimer:2010uk} lead us to expect that,
for a given nuclear matter EoS,
the maximum mass  $M_{\rm max}$
is mostly determined by the central energy density
of the heaviest star
and the speed of sound in the central regions of that star. Specifically,
\beq
M_{\rm max} \overset{?}{=} y(c_{\rm cent}) \ecent^{-1/2} \ .
\label{eqn:M_of_e}
\eeq
The function $y(c_{\rm cent})$ can be obtained from 
Ref.~\cite{Lattimer:2010uk}
(their Eq.~(24) and the associated table). 
To test Eq.~\eqn{eqn:M_of_e}, we follow  
Ref.~\cite{Lattimer:2004sa,Lattimer:2010uk} and plot the maximum
mass $\Mmax$ that a stable star can have,
as a function of the central energy density in that star
$\ecent$. We use the NL3 
nuclear matter 
EoS and
repeat the calculation for a range of quark matter equations of state,
varying the transition density  $\ntrans$ for
the transition to quark matter,
the quark matter speed of sound $\cQM$, and
the energy density discontinuity $\De\ep$ at the transition.
The range of allowed values of $\De\ep$, giving stable hybrid stars
is from zero to $\destab(\ntrans)$.

The results are shown in Fig.~\ref{fig:Mm-ec}, where we use the NL3
nuclear matter EoS and CSS Quark Matter.  The solid curves
show $M_{\rm max}(\ecent)$ according to Eq.~\eqn{eqn:M_of_e}
for $c_{\rm cent}^2=1$, 0.6, and $1/3$.
Nuclear matter equations of state at high density have $c^2$ close to 1,
hence the maximum masses for pure nuclear matter
stars (we show the result for APR \cite{Akmal:1998cf} nuclear matter
as well as NL3 and HLPS)
lie close to the  $c_{\rm cent}^2=1$ line. 

The triangles, squares, and circles show the masses of hybrid stars for
a quark matter EoS with relatively low transition densities:
$\ntrans=1.5\,n_0$ (open symbols)
or  $\ntrans=2\,n_0$ (solid symbols), and for each of these we vary $\cQM$ 
and $\De\ep$.
The low transition density means that these hybrid stars
have large quark matter cores, so the the speed of sound in the central
part of the star is $\cQM$, so the maximum masses should fall on the
lines given by Eq.~\eqn{eqn:M_of_e} with $c_{\rm cent}=\cQM$.
We see that on the whole this is the case.
However, for $\ntrans=2.0\,n_0$ and $\cQMsq=1/3$ (solid triangles)
the points fall slightly below the predicted $\cQMsq=1/3$ line.
This implies that in these stars
the effective speed of sound in the core is even lower
than $1/\sqrt{3}$. However, the speed of sound in the NL3 nuclear matter
mantle at $n\sim 2n_0$ is greater than $1/\sqrt{3}$.
It seems reasonable to argue that
the first-order phase transition, at which
$dp/d \ep = c^2$ vanishes, acts like a ``soft'' region where $c$ is
small \cite{Zdunik:2012dj}, and drags down the average value of $c_{\rm cent}$.
All hybrid stars have such a transition region, but in these stars
the quark matter cores are smaller than in the other cases, 
so the nuclear-quark matter
boundary is closer to the center of the star, and 
by the conjecture of Ref.~\cite{Lattimer:2004sa}
it is expected to play a more important role in determining the maximum mass.

\subsection{How heavy can a hybrid star be?}

Using the CSS parameterization \eqn{eqn:EoSqm1} of the quark matter
EoS, it is possible to get hybrid 
stars that are heavy enough to be consistent with recent
measurements of stars of mass $2\,M_\odot$ 
\cite{Demorest:2010bx,Antoniadis:2013pzd}.
In Fig.~\ref{fig:maxM-QM-params} we show contour plots for 
the maximum masses of hybrid stars as a function of the parameters of the quark
matter EoS, for both the HLPS and NL3 nuclear equations of state.
Fig.~\ref{fig:maxM-QM-params} is a generalization of Fig.~10 of 
Ref.~\cite{Zdunik:2012dj}, 
showing the effect of different nuclear matter EoSs and different
transition densities.

We see that, as expected from  Fig.~\ref{fig:phase-diag-HLPS},
hybrid stars will be heavier if the energy density discontinuity $\De\ep$ 
is smaller (so the gravitational pull of the core does not destabilize the star)
and the speed of sound in quark matter is higher (so the core is stiffer
and can support a heavy star).

For the softer HLPS EoS, whose pure nuclear matter star has a maximum mass of 
2.15\,$M_\odot$ (Table \ref{tab:EoSproperties}), 
hybrid stars can be heavier than the pure nuclear star.
This occurs when the transition density is low, and
the quark core is a large fraction of the star. In the $M(R)$ relation, the
hybrid branch can have a positive $dM/dR$, so its
$M(R)$ curve looks similar to that of a pure quark matter star,
rising to a maximum mass star which is heavier and larger than the 
heaviest pure HLPS star.
For the harder NL3 EoS, whose pure nuclear matter star has a maximum mass of 
2.77\,$M_\odot$, hybrid stars are always lighter than the
heaviest nuclear matter star.

\subsection{The quark matter mass fraction}

For hybrid stars, it is natural to ask how much of the mass of the star
consists of quark matter. We define
the quark matter mass fraction $f_q\equiv M_{\rm core}/M_{\rm star}$.
In  Fig.~\ref{fig:maxM-QM-params}(a), along the thick (blue, $\ntrans= 4 n_0$)
contours of constant $M_{\rm star}$,
$f_q$ varies from about 60\% at low $\cQMsq$ and $\De\ep$
to about 70\% at high  $\cQMsq$ and $\De\ep$. 
If the transition pressure
is lower then we expect the quark fraction to be larger:
along the medium-thickness (green) mass contours
for $\ntrans= 2 n_0$, $f_q$ varies from about 90\% at low $\cQMsq$ and $\De\ep$
to about 96\% at high  $\cQMsq$ and $\De\ep$.

In  Fig.~\ref{fig:maxM-QM-params}(b), along the medium-thickness (green)
mass contours
for $\ntrans= 2 n_0$, $f_q$ varies from about 50\% at low $\cQMsq$ and $\De\ep$
to about 80\% at high  $\cQMsq$ and $\De\ep$. Again, if the transition pressure
is lower then the quark fraction is larger:
along the thin (red) mass contours
for $\ntrans= 1.5 n_0$, $f_q$ varies from about 80\% at low $\cQMsq$ and $\De\ep$
to over 90\% at high  $\cQMsq$ and $\De\ep$.

\section{Conclusions}
\label{sec:conclusions}

We studied hybrid stars where there is a sharp interface between
two phases with different equations of state. We called the two phases
``nuclear matter'' and ``quark matter'', but our conclusions are valid
for any first-order phase transition between two phases with different 
energy densities.

It is already known that in the presence of such a first-order
phase boundary there will be a 
stable connected branch of hybrid stars if the 
energy density discontinuity
$\De\ep$ at the transition is less than a critical value
$\destab$ (Eq.~\eqn{eqn:stability}) which depends 
only on the ratio of pressure to nuclear matter energy density
at the transition. 
We confirmed this result and investigated the conditions for the
occurrence of a disconnected hybrid branch, and the visibility
of the hybrid branches.
To study these properties of hybrid stars we used
a fairly generic parameterization (CSS) of the quark matter EoS 
at densities beyond the transition, and came to the following
conclusions.\\
1) Even if there is no connected hybrid star branch, there may be a
disconnected one if the transition density is low enough
and the speed of sound is high enough so that
the nuclear mantle can be supported by a large enough quark matter core,
and the energy density discontinuity 
is large enough so that a medium-sized core has insufficient pressure
to support the nuclear mantle (see Fig.~\ref{fig:phase-diag-HLPS}).\\
2) The phase diagram for disconnected hybrid star branches is
largely determined by the parameters of the QM EoS,
$\ptrans/\etrans$, $\De\ep/\etrans$, and $\cQMsq$.
It is not very sensitive to
the detailed form of the nuclear matter EoS.\\
3) Even if, according to the standard criterion \eqn{eqn:stability},
a connected hybrid branch is present, it may be so short as to be
very hard to detect by measurements of the mass and radius of the star
(see Fig.~\ref{fig:phase-diag-Mcontours}). When a disconnected branch
is present, the connected branch is either absent or very small.\\
4) We confirmed that the relationship \eqn{eqn:M_of_e}
between the maximum mass of a star and its central energy density, which
was proposed in Ref.~\cite{Lattimer:2010uk}, holds for hybrid stars
(see Fig.~\ref{fig:Mm-ec}).\\
5) We found that it is possible to
get heavy hybrid stars (more than $2\,M_\odot$) for reasonable parameters
of the quark matter EoS. It requires a not-too-high transition density 
($\ntrans\sim 2n_0$), low enough energy density discontinuity
$\De\ep\lesssim 0.5\etrans$, and high enough speed of sound
$\cQMsq \gtrsim 0.4$). It is interesting to note that free quark matter
would have $\cQMsq=1/3$, and a value of $\cQMsq$ that is well above 1/3
is an indication that the quark matter is strongly coupled.
For a stiff nuclear matter EoS such as NL3 it was
somewhat easier to make heavy hybrid stars.
Details are presented in Fig.~\ref{fig:maxM-QM-params}.
Our findings provide a generic formulation of
the results found in previous calculations for specific models
of quark matter, e.g., \cite{Alford:2004pf,Masuda:2012ed},
which found that in the absence of theoretical or 
experimental constraints on the
quark matter EoS, one can fairly easily vary the
unknown parameters of that EoS to obtain heavy hybrid stars.
For now, it seems clear that without theoretical advances
that constrain the form of the quark matter EoS, measurements of gross
features of the $M(R)$ curve such as the maximum mass will not be able to
rule out the presence of quark matter in neutron stars.

Our overarching message is
that the generic CSS parameterization of the quark matter EoS
(also used in Ref.~\cite{Zdunik:2012dj})
provides a general framework
for comparison and empirical testing of models of the quark matter EoS.
Any particular model can be characterized, at least at densities that
are not too far from
the transition, in terms of its values of the parameters $\ptrans/\etrans$, 
$\De\ep/\etrans$, and $\cQMsq$,
and its predictions for hybrid stars will follow from these values,
which determine its
position in the phase diagram (Fig.~\ref{fig:phase-diag-HLPS}).
If the form of the nuclear matter EoS were established then
measurements of the $M(R)$ relation of neutron stars
could be directly expressed as constraints on the values of our 
quark matter EoS parameters.

The CSS parameterization relies on the assumption of a density-independent 
speed of sound, which is a useful starting point
for general comparisons between quark matter models, as well as providing
specific examples of quark matter equations of state that can yield
heavy hybrid stars  (Fig.~\ref{fig:maxM-QM-params}).
If observations of $M(R)$ for heavy stars turned out to be
inconsistent with this parameterization, or if theorists were able to
show that the speed of sound in quark matter has significant density
dependence, then
this approach could be further generalized (at the penalty of introducing more 
parameters) to allow for that, and to allow for mixed or percolated phases as 
noted in Refs.~\cite{Macher:2004vw,Masuda:2012ed}.

Finally, it would be valuable to extend the work reported here
by studying rotating hybrid stars
using the CSS parameterization. It is interesting to note
that a study of a wide range of quark matter EoSs
\cite{Zdunik:2005kh} found that the topology of the hybrid branch
was not affected by rotation.

\section*{Acknowledgments}

We thank David Blaschke, Pawel Haensel, and J\"urgen Schaffner-Bielich
for pointing out relevant earlier work.
This research was supported in part by the 
Offices of Nuclear Physics and High Energy Physics of the
Office of Science of the 
U.S.~Department of Energy under contracts
\#DE-FG02-91ER40628 (for M.G.A.),  
\#DE-FG02-05ER41375 (for M.G.A. and S.H.), 
\#DE-FG02-93ER-40756 (for M.P.), 
 and by the DOE Topical Collaboration 
 ``Neutrinos and Nucleosynthesis in Hot and Dense Matter'', 
 contract \#DE-SC0004955.

\appendix

\section{Constant-Sound-Speed equation of state}
\label{sec:ideal}

Here we briefly recapitulate (see, e.g., Ref.~\cite{Zdunik:2012dj})
the construction of a
thermodynamically consistent equation of state
of the form in Eq.~\eqn{eqn:EoSqm1}
\beq
\varepsilon\left(p\right) = \varepsilon_{0}+\frac{1}{c^2} p \ .
\label{eqn:ep_para}
\eeq
We start by writing the pressure in terms of the chemical potential
\beq
\ba{rcl}
p(\mu_{B}) &= &\dsp A \,\mu_{B}^{1+\beta} - B \ , \\[2ex]
\mu_{B}(p) &=&\dsp   \left(\frac{p+B}{A}\right)^{1/(1+\beta)}
\ .
\ea
\label{eqn:mup}
\eeq
Note that we have introduced an additional parameter $A$ with mass
dimension $3-\be$.
The value of $A$ can
be varied without affecting the energy-pressure relation \eqn{eqn:ep_para}. When constructing a first-order transition from some low-pressure EoS to
a high-pressure EoS of the form \eqn{eqn:ep_para}, we must
choose $A$ so that the pressure is a monotonically
increasing function of $\mu_B$ (i.e. so that the jump in $n_B$ at the transition
is not negative). The derivative with respect to $\mu_B$ yields
\beq
n_{B}(\mu_{B}) =  (1+\beta)\,A \,\mu_{B}^{\beta}
\label{eqn:nmu}
\eeq
and using $p = \mu_B n_B-\ep$, we obtain the energy density
\beq
\ep(\mu_{B})  =  B + \beta\,A\,\mu_{B}^{1+\beta} \ .
\label{eqn:emu}
\eeq
Then Eq.~\eqn{eqn:mup} gives energy density as a function of pressure
\beq
\varepsilon(p)  =  (1+\beta) B + \beta p 
\label{eqn:ep}
\eeq
which is equivalent to Eq.~\eqn{eqn:ep_para} with $1/c^2=\be$ and
$\ep_0= (1+\beta) B $.

\renewcommand{\href}[2]{#2}

\newcommand{\apjl}{Astrophys. J. Lett.\ }
\newcommand{\mnras}{Mon. Not. R. Astron. Soc.\ }
\newcommand{\aap}{Astron. Astrophys.\ }

\bibliographystyle{JHEP_MGA}
\bibliography{hybrid_star} 

\end{document}